\documentclass[12pt]{iopart}

\usepackage{graphicx}

\begin{document}

\title[Anderson localization in randomly-stratified anisotropic media]{Anderson localization and Brewster anomaly of electromagnetic waves in randomly-stratified anisotropic media}

\author{Kihong Kim$^{1,2}$ and Seulong Kim$^1$}

\address{$^1$ Department of Energy Systems
Research and Department of Physics, Ajou University, Suwon 16499}
\address{$^2$ School of Physics, Korea Institute for Advanced Study, Seoul 02455, Korea}

\ead{khkim@ajou.ac.kr}

\begin{abstract}
Anderson localization of $p$-polarized waves and the Brewster anomaly phenomenon, which is the delocalization of $p$-polarized waves at a special incident angle,
in randomly-stratified anisotropic media are studied theoretically for two different random models. In the first model, the random parts of the transverse and longitudinal components of the dielectric tensor, between which
the longitudinal component is the one in the stratification direction, are assumed to be uncorrelated, while, in the second model,
they are proportional to each other.
We calculate the localization length in a precise way using the invariant imbedding method.
From analytical considerations, we provide an interpretation of the Brewster anomaly as a phenomenon arising
when the wave impedance is effectively uniform. Similarly, the ordinary Brewster effect is interpreted as an impedance matching phenomenon.
We derive the existence condition for the Brewster anomaly and concise analytical expressions for the localization length, which are accurate in the weak disorder regime.
We find that the Brewster anomaly can arise only when disorder is sufficiently weak and only
in the second model with a positive ratio of the random parts. The incident angle at which the anomaly occurs
depends sensitively on the ratio of the random parts and the average values of the tensor components. In the cases where the critical angle of total
reflection exists, the angle at which the anomaly occurs can be either bigger or smaller than the critical angle.
When the transverse and longitudinal components are
uncorrelated, localization is dominated by the the transverse component at small incident angles.
When only the longitudinal component is random, the localization length diverges as $\theta^{-4}$ as the incident angle $\theta$ goes to zero
and is also argued to diverge for all $\theta$ in the strong disorder limit.
\end{abstract}

\maketitle

%%%%%%%%%%%%%%%%%%%%%%%%%%  body  %%%%%%%%%%%%%%%%%%%%%%%%%%

\section{Introduction}

Even though it has been studied extensively for over half a century, Anderson localization of quantum particles and classical waves continues
to attract the interest of many researchers \cite{pwa,mod,gre,seg,seg2,seg3,jeon,kk0}.
%In this paper, we are interested in the localization of electromagnetic waves in randomly-stratified anisotropic media,
%which can be encountered frequently among both naturally-occurring media and fabricated metamaterials \cite{lag,jor1,jor2,gev}.
We focus especially on the unique phenomenon called Brewster anomaly (BA), which is the delocalization of $p$-polarized electromagnetic waves
in randomly-stratified media
at a special incident angle \cite{sipe,lee,mog,rey,kiv,lis,aga}. Understanding the mechanism of this phenomenon in anisotropic media,
which can be encountered frequently among both naturally-occurring media and fabricated metamaterials,
is crucial in the development of polarization-insensitive reflectors and polarization-sensitive optical devices, as well as in understanding
some bio-optical properties \cite{lag,jor1,jor2,gev,iwa,feller,mei,upa}.

Since the discovery of the BA by Sipe {\it et al.} \cite{sipe}, many authors have discussed different aspects of this phenomenon.
Jordan {\it et al.} have studied the BA occurring in randomly-layered anisotropic media consisting of alternating isotropic-uniaxial
media using numerical calculations based on the transfer matrix method \cite{jor1}. They have found that in the cases called mixed stacks,
the BA is suppressed and does not occur, while it can occur in the cases called binary stacks.
A similar model based on alternating metamaterial-uniaxial randomly layered stacks has been studied by del Barco {\it et al.}
using the transfer matrix method, where the BA is again found to be suppressed \cite{gev}.

In this paper, we will present a unique perspective on the BA that
it is a phenomenon arising
when the effective wave impedance is uniform and non-random, which is made possible only in weakly-disordered media.
By a similar argument, we will also argue that the ordinary Brewster effect arises when the wave impedance is completely matched
throughout the space.
Using an analytical method based on the invariant imbedding theory \cite{kly,kim}, we will derive precise conditions for the occurrence of the BA in
randomly-stratified anisotropic media
and derive concise analytical expressions for the localization length in the weak disorder regime for two different random models.
These results will be compared with more accurate numerical results obtained using the invariant imbedding method and
also with the previous results obtained for randomly-layered anisotropic media \cite{jor1,gev}.
In addition, we will derive some interesting properties of localization in anisotropic media from general analytical considerations.

\section{Model}

We consider a random uniaxial medium, the dielectric permittivity tensor of which is diagonalized in the coordinate
system $(x,y,z)$ and is written as
\begin{eqnarray}
\epsilon=\pmatrix{\epsilon_\perp & 0 & 0\cr 0 & \epsilon_\perp &0 \cr 0 & 0 & \epsilon_\parallel}.
\end{eqnarray}
The medium is stratified along the $z$ axis and the transverse and longitudinal tensor components, $\epsilon_\perp$ and $\epsilon_\parallel$, are random functions of $z$ only.
Plane electromagnetic waves of frequency $\omega$ and vacuum wave number $k_0$ ($=\omega/c$) are
assumed to propagate in the $xz$ plane. Then the wave equations for the $s$- and $p$-polarized waves are completely
decoupled. In this paper, we are only interested in the propagation
of $p$ waves, for which the $y$ component of the magnetic field satisfies
\begin{eqnarray}
{H_y}^{\prime\prime}-\frac{{\epsilon_\perp}^\prime}{\epsilon_\perp}{H_y}^\prime+\left({k_0}^2\epsilon_\perp
-q^2\frac{\epsilon_\perp}{\epsilon_\parallel}\right)H_y=0,\label{eq:pw}
\end{eqnarray}
where $q$ is the $x$ component of the wave vector and a prime denotes a differentiation with respect to $z$.

We assume that an inhomogeneous anisotropic medium
of thickness $L$ lies in $0\le z\le L$ and the waves are incident obliquely from a uniform dielectric region ($z>L$)
and transmitted to another uniform dielectric region ($z<0$). The incident and transmitted regions are filled with ordinary isotropic
media of the same kind, where $\epsilon$ ($=\epsilon_1$) is a scalar quantity. When $\theta$ is
the angle of incidence, $q$ is equal to $k\sin\theta$, where $k=k_0\sqrt{\epsilon_1}$.
From now on, we will assume that {\it $\epsilon_\perp$ and $\epsilon_\parallel$ are always normalized by $\epsilon_1$} to
simplify the notations, unless otherwise explicitly stated.

We consider two different random models. In Model I,
we assume that
$\epsilon_\perp$ and $\epsilon_\parallel$ are {\it independent} random functions
of $z$ and satisfy
\begin{eqnarray}
\epsilon_\perp=a+\delta\epsilon_\perp(z),
~~\epsilon_\parallel=b+\delta\epsilon_\parallel(z),
\label{eq:weak}
\end{eqnarray}
where $a$ and $b$ are the disorder-averaged values of $\epsilon_\perp$ and $\epsilon_\parallel$
and $\delta\epsilon_\perp(z)$ and $\delta\epsilon_\parallel(z)$ are Gaussian random functions satisfying
\begin{eqnarray}
&&\langle \delta\epsilon_\perp(z)\delta\epsilon_\perp(z^\prime)\rangle={\tilde g_\perp}\delta(z-z^\prime),~~\langle\delta\epsilon_\perp(z)\rangle=0,\nonumber\\
&&\langle \delta\epsilon_\parallel(z)\delta\epsilon_\parallel(z^\prime)\rangle={\tilde g_\parallel}\delta(z-z^\prime),~~\langle\delta\epsilon_\parallel(z)\rangle=0.
\end{eqnarray}
The notation $\langle\cdots\rangle$ denotes averaging over disorder and $\tilde g_\perp$ and $\tilde g_\parallel$ are
independent parameters characterizing the strength of disorder.
On the other hand, in Model II, we consider the situation where the random components
$\delta\epsilon_\perp(z)$ and $\delta\epsilon_\parallel(z)$ are not independent, but proportional to each other such that
\begin{equation}
\delta\epsilon_\parallel(z)=f\delta\epsilon_\perp(z),
\end{equation}
where $f$ is a real constant.

\section{Invariant imbedding method}

Since the BA occurs only for $p$ waves, we focus on that case here.
We consider a $p$ wave of unit magnitude incident
on the anisotropic medium.
Using the invariant imbedding method and starting from Eq.~(\ref{eq:pw}), we derive exact differential
equations satisfied by the reflection
and transmission coefficients, $r$ and $t$:
\begin{eqnarray}
&&\frac{1}{ip}\frac{dr}{dl}=2\epsilon_\perp r+\frac{1}{2}\left(\sec^2\theta-{\epsilon}_\perp-\frac{\tan^2\theta}{\epsilon_\parallel}\right)\left(1+r\right)^2,
\nonumber\\
&&\frac{1}{ip}\frac{dt}{dl}=\epsilon_\perp t+\frac{1}{2}\left(\sec^2\theta-{\epsilon}_\perp-\frac{\tan^2\theta}{\epsilon_\parallel}\right)\left(1+r\right)t,
\label{eq:pp}
\end{eqnarray}
where $p$ ($=k\cos\theta$) is the negative $z$ component of the wave vector
in the incident and transmitted regions.
We use Eq.~(\ref{eq:pp})
for the precise numerical calculation of the localization length $\xi$ defined by
\begin{eqnarray}
\xi=-\lim_{L\rightarrow\infty}\left(\frac{L}{\langle \ln T\rangle}\right),
\end{eqnarray}
where $T$ is the transmittance given by $T=\vert t\vert^2$.

The invariant imbedding equations for $r$ and $t$, Eq.~(\ref{eq:pp}), are stochastic differential equations with random coefficients.
In order to deal with the random term $\epsilon_\parallel$ appearing in the denominators of the coefficients in Eq.~(\ref{eq:pp})
by using known methods, we assume that the disorder in $\epsilon_\parallel$ is sufficiently weak
so that
\begin{eqnarray}
\frac{1}{{\epsilon}_\parallel}=\frac{1}{b+\delta\epsilon_\parallel}\approx\frac{1}{b}-\frac{\delta\epsilon_\parallel}{b^2}.
\label{eq:app}
\end{eqnarray}
We point out that this is the only approximation used in the present work.
In contrast, the disorder in $\epsilon_\perp$ can be of arbitrary strength.
From the general considerations presented in Sec.~\ref{sec:gc}, we will show that
the BA can occur only when disorder is sufficiently weak. Therefore, the condition given in Eq.~(\ref{eq:app})
is one of the necessary conditions for the existence of the BA, rather than an approximation.

\subsection{Model I}

In order to obtain the localization length, we need to compute the
average $\langle \ln T(L)\rangle$ in the $L\rightarrow \infty$ limit.
The $\it nonrandom$ differential equation satisfied by $\langle \ln
T\rangle$ can be obtained using the second of Eq.~(\ref{eq:pp}), Eq.~(\ref{eq:app}) and Novikov's formula \cite{nov}
and takes the form
%\begin{widetext}
\begin{eqnarray}
-{1\over k}{{d\langle \ln T\rangle}\over{dl}}=
C_1+{\rm Re}\left[
\left(iC_0-2C_2\right)Z_{1}+C_1Z_{2}\right],
\label{eq:w3}
\end{eqnarray}
%\end{widetext}
where $Z_n$ ($n=1,2$) is equal to $\langle r^n\rangle$ and
the parameters $C_0$, $C_1$ and $C_2$ are defined by
\begin{eqnarray}
&&C_0=\left(a+\frac{\tan^2\theta}{b}-\sec^2\theta\right)\cos\theta,~~
C_1=g_\perp\cos^2\theta+g_\parallel\frac{\tan^2\theta\sin^2\theta}{b^4},\nonumber\\
&&C_2=g_\perp\cos^2\theta-g_\parallel\frac{\tan^2\theta\sin^2\theta}{b^4}.
\label{eq:zz2}
\end{eqnarray}
The $\it dimensionless$ disorder parameters $g_\perp$ and $g_\parallel$ are given by
\begin{equation}
g_\perp=\frac{{\tilde g_\perp k}}{4},~~g_\parallel=\frac{{\tilde g_\parallel k}}{4}.
\end{equation}
In Model I, the random terms
$\delta\epsilon_\perp$ and $\delta\epsilon_\parallel$ are {\it uncorrelated}. This fact has played an important role in deriving
Eq.~(\ref{eq:w3}).
In the $l\rightarrow \infty$ limit, the left-hand side of Eq.~(\ref{eq:w3}) approaches asymptotically to a constant equal to
$(k\xi)^{-1}$.

To calculate $Z_1$ and $Z_2$ for use in Eq.~(\ref{eq:w3}), we derive an infinite number of coupled nonrandom differential
equations satisfied by $Z_{n}$, where $n$ is an arbitrary nonnegative integer,
using the first of Eq.~(\ref{eq:pp}) and Novikov's formula.
These equations turn out to take the form
%\begin{widetext}
\begin{eqnarray}
{1\over k}{{dZ_{n}}\over{dl}}&=& in\cos\theta\left(a-\frac{\tan^2\theta}{b}+\sec^2\theta\right)Z_{n}
-\frac{i}{2}nC_0\left(Z_{n+1}+Z_{n-1}\right)-3n^2C_1Z_{n}\nonumber\\
&&+\left(2n+1\right)n C_2 Z_{n+1}+\left(2n-1\right)n C_2 Z_{n-1}
\nonumber\\&&-\frac{1}{2}n(n+1)C_1 Z_{n+2}-\frac{1}{2}n(n-1)C_1 Z_{n-2}.
\label{eq:w1}
\end{eqnarray}
%\end{widetext}
The initial conditions for $Z_{n}$'s are $Z_{0}=1$ and
$Z_{n}(l=0)=0$ for $n>0$. In the $l\rightarrow \infty$ limit,
the left-hand sides of these equations vanish and we obtain an infinite
number of coupled $\it algebraic$ equations, which are much easier to solve numerically than
the coupled differential equations.
The moments
$Z_{n}$ with $n > 0$ are coupled to one another
and their magnitudes decrease rapidly as $n$ increases.
Based on this observation, we solve these equations numerically by a
systematic truncation method \cite{kim1}.

\subsection{Model II}

In Model II, $\delta\epsilon_\perp(z)$ and $\delta\epsilon_\parallel(z)$ are not independent, but proportional to each other.
This condition leads to completely different equations for $Z_{n}$ and $\langle \ln T\rangle$ for $p$ waves.
The equation for $Z_{n}$ in this case is written as
%\begin{widetext}
\begin{eqnarray}
{1\over k}{{dZ_{n}}\over{dl}}&=& in\cos\theta\left(a-\frac{\tan^2\theta}{b}+\sec^2\theta\right)Z_{n}-\frac{i}{2}nC_0\left(Z_{n+1}+Z_{n-1}\right)\nonumber\\
&&-g_\perp\left[3\left(1+\frac{f^2}{b^4}\tan^4\theta\right)\cos^2\theta
+2 \frac{f}{b^2}\sin^2\theta\right] n^2 Z_{n}\nonumber\\
&&+\left(2n+1\right)n D_2 Z_{n+1}
+\left(2n-1\right)n D_2 Z_{n-1}\nonumber\\
&&-\frac{1}{2}n(n+1)D_1 Z_{n+2}-\frac{1}{2}n(n-1)D_1 Z_{n-2},
\label{eq:w1b}
\end{eqnarray}
%\end{widetext}
where the parameters $D_1$ and $D_2$ are defined by
\begin{eqnarray}
D_1=g_\perp\left(1-\frac{f}{b^2}\tan^2\theta\right)^2\cos^2\theta,~~
D_2=g_\perp\left(1-\frac{f^2}{b^4}\tan^4\theta\right)\cos^2\theta.
\label{eq:zz22}
\end{eqnarray}
The equation for the localization length takes the form
%\begin{widetext}
\begin{eqnarray}
-{1\over k}{{d\langle \ln T\rangle}\over{dl}}=
D_1+{\rm Re}\left[
\left(iC_0-2D_2\right)Z_{1}+D_1Z_{2}\right].
\label{eq:w44}
\end{eqnarray}
%\end{widetext}

\section{General considerations on the existence condition of the Brewster anomaly
and the properties of localization}
\label{sec:gc}

\subsection{Argument based on the impedance matching condition}

There is a very simple interpretation of the BA phenomenon, which has never been, to our knowledge,
advocated before. Based on this interpretation,
it is possible to explain both the Brewster effect and the BA phenomenon in a
unified way. Furthermore, we can deduce some interesting properties of localization
in anisotropic media.
We begin by rewriting the wave equation, Eq.~(\ref{eq:pw}), in the following equivalent form:
\begin{eqnarray}
\left(\frac{{H_y}^{\prime}}{\epsilon_\perp}\right)^\prime
+p^2\epsilon_\perp\eta^2H_y,
\label{eq:imp}
\end{eqnarray}
where $\eta$ is defined by
\begin{eqnarray}
\eta^2=\frac{\epsilon_\parallel-\sin^2\theta}{\epsilon_\perp \epsilon_\parallel \cos^2\theta}.
\label{eq:imp2}
\end{eqnarray}
In these expressions, we remind again that $\epsilon_\perp$ and $\epsilon_\parallel$ are quantities normalized by $\epsilon_1$. Therefore,
in the incident and transmitted regions where $\epsilon_\perp=\epsilon_\parallel=1$, $\eta$ is equal to 1 for all $\theta$.
We notice that the wave equation written in the above form looks identical to that for $p$ waves propagating normally in a
medium with the {\it wave impedance} given by $\eta(z)$.

Before discussing the BA, it is instructive to examine the ordinary Brewster effect from the viewpoint of impedance matching.
It is well-known that if the entire medium has a uniform impedance, waves are completely transmitted without any backward reflection.
In our case, the uniform impedance condition requires $\eta$ to be equal to 1 in the entire slab.
From Eq.~(\ref{eq:imp2}), it is straightforward to derive the incident angle $\theta_b$, which is nothing but
the ordinary Brewster angle, for total transmission of $p$ waves. We obtain
\begin{eqnarray}
\tan^2\theta_b=\frac{\epsilon_\parallel \left(\epsilon_\perp -1\right)}{\epsilon_\parallel-1}.
\label{eq:ba}
\end{eqnarray}
Obviously, the right-hand side of the above equation has to be positive for $\theta_b$ to exist.
The same result has been obtained long ago by other authors \cite{lek,shu}.
In isotropic media, we have $\epsilon_\perp=\epsilon_\parallel$ ($=\epsilon$). Then we reduce Eq.~(\ref{eq:ba}) to the
well-known expression for the Brewster angle, $\tan\theta_b=\sqrt{\epsilon}$.

The BA is a delocalization phenomenon arising
at a special incident angle, $\theta_B$, when $\epsilon_\perp$ and $\epsilon_\parallel$ are random functions of $z$.
In order for delocalization to occur, the impedance $\eta$ needs to be either a real constant or a real-valued {\it nonrandom} function of $z$.
From the functional form of Eq.~(\ref{eq:imp2}), we find that this cannot be
realized if $\epsilon_\perp$ and $\epsilon_\parallel$ are random functions
of arbitrary strength of disorder. For sufficiently weak disorder, however, we substitute Eq.~(\ref{eq:weak}) into Eq.~(\ref{eq:imp2})
and use the Taylor expansion to transform it to
\begin{eqnarray}
\eta^2\approx \frac{b-\sin^2\theta}{ab\cos^2\theta}+
\frac{\left(a\sin^2\theta\right)\delta\epsilon_\parallel-b\left(b-\sin^2\theta\right)\delta\epsilon_\perp}{a^2b^2\cos^2\theta},
\label{eq:appr}
\end{eqnarray}
to the first order in $\delta\epsilon_\perp$ and $\delta\epsilon_\parallel$.
The only nontrivial possibility for $\eta$ to be nonrandom in Eq.~(\ref{eq:appr}) is when
\begin{equation}
\left(a\sin^2\theta\right)\delta\epsilon_\parallel=b\left(b-\sin^2\theta\right)\delta\epsilon_\perp,
\end{equation}
while both $\delta\epsilon_\perp$ and $\delta\epsilon_\parallel$ are nonzero. If $\delta\epsilon_\parallel$ is zero and
$b$ is equal to $\sin^2\theta$, $\eta$ becomes zero and the wave does not propagate.
Therefore $\delta\epsilon_\perp$ and $\delta\epsilon_\parallel$ have to be proportional to each other, as in our Model II,
where $\delta\epsilon_\parallel=f\delta\epsilon_\perp$. We finally obtain
\begin{equation}
\sin\theta_B=\left(\frac{b^2}{b+af}\right)^{1/2},
\label{eq:angle1}
\end{equation}
where $(b+af)$ has to be positive and bigger than $b^2$ to have a solution for $\theta_B$.
In the case of isotropic media with $f=1$ and $b=a$, this reduces to the well-known result, $\sin\theta_B=\sqrt{a/2}$,
derived originally by Sipe {\it et al.} \cite{sipe}.

By substituting Eq.~(\ref{eq:angle1}) into Eq.~(\ref{eq:appr}), we obtain the effective wave impedance
when a $p$ wave is incident at $\theta_B$ given by
\begin{equation}
\eta_B = \left(\frac{f}{b+af-b^2}\right)^{1/2}.
\end{equation}
In order to have a propagating wave, the wave impedance needs to be real, which gives an additional
constraint such that $f$ has to be positive. In other words, the random functions $\delta\epsilon_\perp(z)$ and $\delta\epsilon_\parallel(z)$
need to be always of the same sign.
When the constraints $b+af>b^2$ and $f>0$ are satisfied, the angle $\theta_B$ is well-defined and the impedance $\eta_B$
is a positive real constant, which is not generally equal to 1.
Since $\eta_B$ is not matched to that of the incident region in general,
the wave incident at $\theta_B$ on a randomly-stratified slab of finite thickness is partially
reflected and the disorder-averaged transmittance is smaller than 1 and depends on the thickness.

\begin{figure}[htbp]
\centering
\includegraphics[width=10cm]{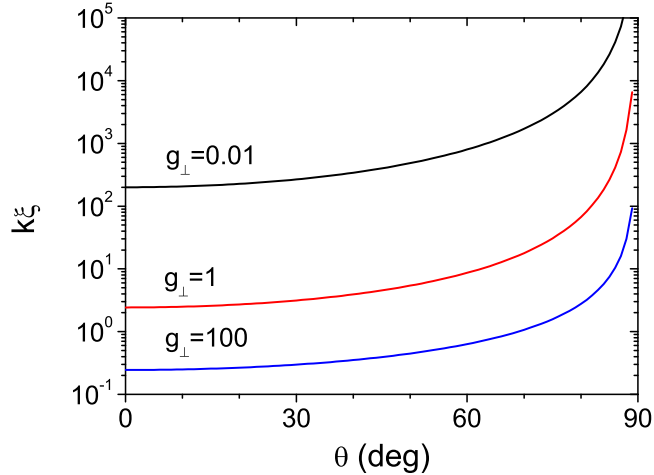}
\caption{Normalized localization length, $k\xi$, versus incident angle, $\theta$, for $p$ waves in Model I, when $a=2$, $b=1$, $g_\parallel=0$
and $g_\perp=0.01$, 1, 100. $\xi$ diverges as $\theta$ approaches $90^\circ$.}
\label{fig1}
\end{figure}

From the simple form of $\eta$ in Eq.~(\ref{eq:imp2}), we can also deduce several additional properties of localization
in anisotropic media. If the incident angle $\theta$ is zero, then the dependence on $\epsilon_\parallel$ disappears in Eq.~(\ref{eq:imp2}),
which reduces to $\eta^2={\epsilon_\perp}^{-1}$. This implies that if $\epsilon_\perp$ is nonrandom, Anderson localization does
not occur at $\theta=0$ for any random function $\epsilon_\parallel$ and the localization length $\xi$ diverges.
In addition, when $\theta$ is sufficiently close to zero, we find from Eq.~(\ref{eq:appr}) that
the random term $\delta\epsilon_\parallel$ has a coefficient proportional to $\theta^2$, which suggests that
the strength of the $\epsilon_\parallel$ disorder, $g_\parallel$, will always appear as multiplied by $\theta^4$ in this regime.
In later sections, we will present an analytical formula and numerical results showing that,
in the presence of only the $\epsilon_\parallel$ disorder, $\xi$ is
indeed proportional to $(g_\parallel \theta^4)^{-1}$ when $\theta$ is sufficiently small.
Another observation we make about the $\theta=0$ case is that if $\epsilon_\perp$ is equal to 1, that is, if
the transverse tensor component is matched to the permittivity of the incident region,
then the impedance is 1 in all regions of space and therefore the transmission has to be perfect regardless of the form of $\epsilon_\parallel$.
We notice that this behavior shows close similarity to the Klein tunneling of massless Dirac electrons entering a
random scalar potential barrier normally \cite{kk0,zzw,bli,zha}, where the scalar potential plays a similar role as $\epsilon_\parallel$.

Next, we consider the situation where the longitudinal component $\epsilon_\parallel$ is very
strongly disordered, while $\epsilon_\perp$ is nonrandom. Then, in the numerator of Eq.~(\ref{eq:imp2}), $\epsilon_\parallel$ dominates the $\sin^2\theta$ term with high probability and
we get $\eta^2\approx (\epsilon_\perp \cos^2\theta)^{-1}$, which is nonrandom. Therefore the localization length has to diverge for all $\theta$
as $g_\parallel$ goes to infinity, if $\epsilon_\perp$ is positive and nonrandom. This implies that the dependence of $\xi$ on $g_\parallel$
is non-monotonic: as $g_\parallel$ increases from zero to infinity, $\xi$ initially decreases, then increases to infinity. This behavior is again similar to that obtained for massless Dirac electrons in a one-dimensional
random scalar potential \cite{kk0,chan}. Our numerical method described in the previous section relies on the assumption that the disorder in $\epsilon_\parallel$
is sufficiently weak, and therefore it cannot be used to study the limit $g_\parallel\rightarrow\infty$. However, we can use a method based on the formula of differentiation derived by Shapiro and Loginov \cite{shap} to study Anderson localization for arbitrarily strong disorder. This approach is beyond the scope of this paper and will
be presented in a future publication.

Finally, we consider the case where $\epsilon_\parallel$ is equal to 1.
Then the expression for the impedance reduces to $\eta^2={\epsilon_\perp}^{-1}$,
which is independent of $\theta$ for any functional form of $\epsilon_\perp$. In this case, if there is only
one scattering interface, the transmission is independent of the incident angle.
However, if there are more than one interfaces such as in a uniform slab of finite thickness or in the case with inhomogeneous $\epsilon_\perp(z)$,
then the interference of multiply scattered waves will occur. This effect depends on $p$ ($=k\cos\theta$) and $\epsilon_\perp(z)$, therefore the
transmission and other characteristics depend on $\theta$ in general.
As an example, we show in Fig.~1 the localization length as a function of $\theta$ calculated using the invariant imbedding method when $\epsilon_\parallel=1$,
$a=2$ and $g_\perp=0.01$, 1, 100. We remind that when only $\epsilon_\perp$ is random, our method can be applied to any large value of $g_\perp$
and the results shown here are exact. When $g_\perp$ is smaller than 1, $\xi$ is very accurately given by
\begin{equation}
k\xi=\frac{a}{g_\perp\cos^2\theta},
\end{equation}
which is a special case of Eq.~(\ref{eq:af1}) to be derived in the next section.
We notice that the localization length has a strong $\theta$ dependence and diverges as $\theta$ approaches $90^\circ$.
This divergence was pointed out previously by Jordan {\it el al.}, who studied an alternating isotropic-uniaxial random
layered medium using the transfer matrix method \cite{jor1}. However, the behavior of their data shown in Fig.~4(a)
of Ref.~\cite{jor1} is markedly different from ours in that, in their case, as $\theta$ increases from zero, $\xi$ remains almost constant up to
$\theta\sim 60^\circ$, and then increases sharply to infinity as $\theta$ approaches $90^\circ$.
Whether this difference is due to the difference in the models used or some other reason remains to be investigated.

\subsection{Argument based on the Fresnel formula}

Equivalently, we can derive the existence condition of
the BA using the Fresnel formula. We consider our medium as consisting of a large number of
very thin layers. The reflection coefficient between two neighboring layers is written as
\begin{eqnarray}
r=\frac{p/a-p^\prime/a^\prime}{p/a+p^\prime/a^\prime},
\end{eqnarray}
where $p$ ($p^\prime$) is the $z$ component of the wave vector in the first (second) layer
with the parameters $a$ and $b$ ($a^\prime$ and $b^\prime$). $p$ satisfies $p^2=k^2a-q^2a/b$ in uniaxial media.
We suppose that the wave is delocalized at an
incident angle $\theta_B$.
In order for delocalization to occur, the random variation of $a$ and $b$ should not cause any reflection,
and therefore we have the no-reflection condition, $p/a=p^\prime/a^\prime$. We write $a^\prime$ and $b^\prime$ as $a^\prime=a+\delta a$ and
$b^\prime=b+\delta b$, with $\delta a$ and $\delta b$ as small quantities. Substituting these into $p/a=p^\prime/a^\prime$
and using the Taylor expansion, we obtain
\begin{equation}
\left(1-\frac{\sin^2\theta_B}{b}\right)\delta a=\left(\frac{a}{b^2}\sin^2\theta_B\right)\delta b,
\end{equation}
which implies that $\delta b$ has to be proportional to $\delta a $.
Therefore, only Model II can show the BA.
If we define $\delta b=f\delta a$, the condition for the BA becomes
identical to Eq.~(\ref{eq:angle1}).

The same conclusions can be deduced from the expressions for the localization length, Eqs.~(\ref{eq:w3}) and (\ref{eq:w44}).
In one dimension, waves are localized in the presence of even an infinitesimally weak
randomness, except for in some special cases. The fact that a $p$ wave is delocalized at $\theta=\theta_B$ implies that
disorder does not play any role in the wave propagation process and the reflection coefficient $r$ is the same as the value
in the absence of disorder, $r_0$, given by
\begin{equation}
r_0=\frac{\sqrt{a}\cos\theta-\left[1-\left(\sin^2\theta\right)/b\right]^{1/2}}{\sqrt{a}\cos\theta+\left[1-\left(\sin^2\theta\right)/b\right]^{1/2}}.
\end{equation}
After substituting $Z_1=r_0$ and $Z_2=r_0^2$ into Eq.~(\ref{eq:w44}), we find that
the right-hand side of Eq.~(\ref{eq:w44}) vanishes and $\xi$ diverges only when
\begin{equation}
\left(r_0-1\right)^2=\frac{f}{b^2}\left(r_0+1\right)^2\tan^2\theta,
\end{equation}
from which
we conclude that only Model II with $f>0$ can display the BA.

\section{Analytical expressions for the localization length in the weak disorder regime}

Starting from Eqs.~(\ref{eq:w3}), (\ref{eq:w1}), (\ref{eq:w1b}) and (\ref{eq:w44}), it is
possible to derive analytical expressions for the localization length in the weak disorder limit.
We write $r$ as $r=r_0+\delta r$. From numerical calculations, we have verified that
$\langle \delta r\rangle$ and $\langle \left(\delta r\right)
^2 \rangle$ are of the first order in disorder, while
$\langle \left(\delta r\right)^3 \rangle$ is of the second order, except at incident angles close to the critical angle
for total internal reflection. From this consideration, we substitute
\begin{eqnarray}
&&Z_1=r_0+\langle \delta r\rangle,~Z_2=r_0^2+2r_0\langle \delta r\rangle+\langle \left(\delta r\right)
^2 \rangle,\nonumber\\
&&Z_3\approx r_0^3+3r_0^2\langle \delta r\rangle+3r_0\langle \left(\delta r\right)^2 \rangle
\end{eqnarray}
into Eq.~(\ref{eq:w1}) in the $l\rightarrow\infty$ limit when $n=1$ and 2
and obtain two coupled equations for $\langle \delta r\rangle$ and $\langle \left(\delta r\right)^2 \rangle$.
We solve them analytically and substitute the results into Eq.~(\ref{eq:w3})
to the leading order in the disorder parameters. The final expression for the localization length for Model I is
\begin{eqnarray}
\frac{1}{k\xi}=2\sqrt{w}\Theta(w)
+g_\perp\frac{b-\sin^2\theta}{ab}+g_\parallel\frac{a\sin^4\theta}{b^3\left(b-\sin^2\theta\right)},
\label{eq:af1}
\end{eqnarray}
where
\begin{equation}
w=a\left(\frac{\sin^2\theta}{b}-1\right)
\end{equation}
and $\Theta$ is the step function, $\Theta(x)=1$ for $x>0$ and 0 for $x<0$.
Similarly, we obtain the localization length for Model II as
\begin{eqnarray}
\frac{1}{k\xi}=2\sqrt{w}\Theta(w)
+g_\perp\frac{\left[b\left(b-\sin^2\theta\right)-fa\sin^2\theta\right]^2}{ab^3\left(b-\sin^2\theta\right)}.
\label{eq:af2}
\end{eqnarray}
We have found numerically that both of these equations are quite accurate when the disorder parameters are
sufficiently small, except near the region where $w=0$. In the isotropic case with $f=1$ and $b=a$, the second term of Eq.~(\ref{eq:af2}) reduces to
\begin{equation}
\frac{1}{k\xi}=g_\perp\frac{(a-2\sin^2\theta)^2}{a^2(a-\sin^2\theta)}
\end{equation}
derived previously by Sipe {\it et al.} \cite{sipe}.

\section{Numerical results}
\label{sec:nr}

\begin{figure}[htbp]
\centering
\includegraphics[width=10cm]{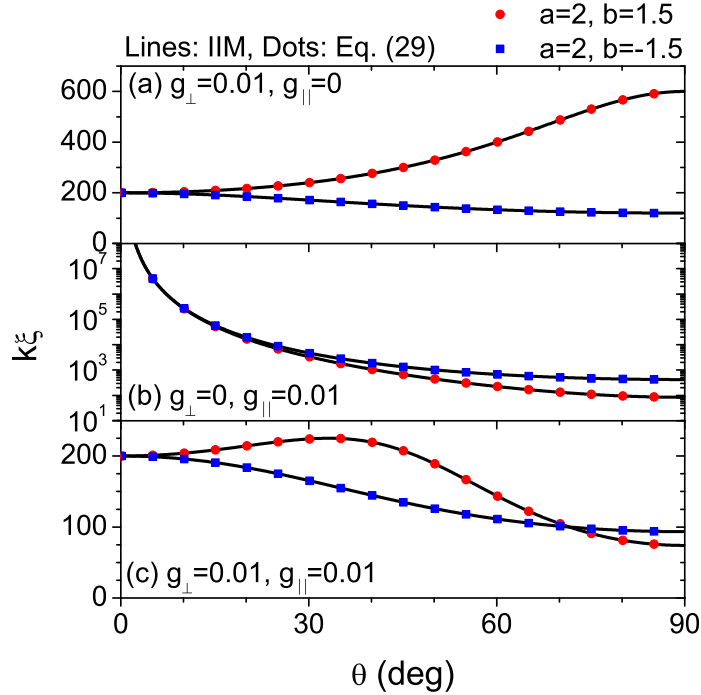}
\caption{Normalized localization length, $k\xi$, versus incident angle, $\theta$, for $p$ waves in Model I, when $a=2$, $b=\pm 1.5$
and (a) $g_\perp=0.01$, $g_\parallel=0$, (b) $g_\perp=0$, $g_\parallel=0.01$, and (c) $g_\perp=g_\parallel=0.01$. The numerical results obtained using the invariant imbedding method are compared with those obtained from the analytical formula, Eq.~(\ref{eq:af1}), which are designated by dots.}
\label{fig2}
\end{figure}

In Fig.~2, we show the normalized localization length, $k\xi$, as a function of the incident angle for Model I, when
$a=2$ and $b=\pm 1.5$. We note that the case with $a>0$ and $b<0$ corresponds to a type I hyperbolic medium \cite{kiv2}.
We consider three cases, where only $\epsilon_\perp$ is random,
only $\epsilon_\parallel$ is random and both $\epsilon_\perp$ and $\epsilon_\parallel$ are
random. In the first case, $\xi$ increases (decreases) monotonically as $\theta$
increases when $b=1.5$ ($b=-1.5$), while, in the second case, it diverges at $\theta=0$ and
decreases monotonically as $\theta$ increases for both $b=\pm 1.5$. The third
case is a combination of the first two cases.

These behaviors can be readily understood from the form of the function $U$ defined by
\begin{eqnarray}
U = 1-\epsilon_\perp+\frac{\epsilon_\perp}{\epsilon_\parallel}\sin^2\theta\approx 1-a-\delta\epsilon_\perp+\left(\frac{a}{b}+\frac{1}{b}\delta\epsilon_\perp-\frac{a}{b^2}\delta\epsilon_\parallel\right)\sin^2\theta,
\label{eq:f2}
\end{eqnarray}
which, in the equivalent Schr\"odinger equation, plays the role of $V(z)/E$,
where $V(z)$ is the potential and $E$ is the energy of an incident quantum particle.
In the case of Fig.~2(a), we find that the strength of the $\delta\epsilon_\perp$ term decreases (increases) monotonically
as $\theta$ increases when $b=1.5$ ($b=-1.5$), in consistence with the behavior of $\xi$.
We notice that if $b=1$, the $\delta\epsilon_\perp$ term will vanish and $\xi$ will diverge, as $\theta$ approaches $90^\circ$.
This case corresponds to that shown in Fig.~4(a) of Ref.~\cite{jor1}, where the longitudinal component of the refractive index is uniform
and matched to that of the surrounding medium.
In the case of Fig.~2(b), the strength of the $\delta\epsilon_\parallel$ term increases from zero monotonically
as $\theta$ increases from zero, regardless of the sign of $b$, which is again in consistence with the behavior of $\xi$.
When only $\epsilon_\parallel$ is random, all normally incident waves are delocalized.
We find that localization is dominated by the randomness of $\epsilon_\perp$ at small incident angles.
The nonmonotonic behavior of $\xi$ shown in Fig.~2(c) when $a=2$ and $b=1.5$ is qualitatively similar to that
shown in Fig.~2(b) of Ref.~\cite{jor1}. We point out that the system called mixed stack in Ref.~\cite{jor1} corresponds to Model I and will not
show the BA, while that called binary stack can show it.

\begin{figure}[htbp]
\centering
\includegraphics[width=10cm]{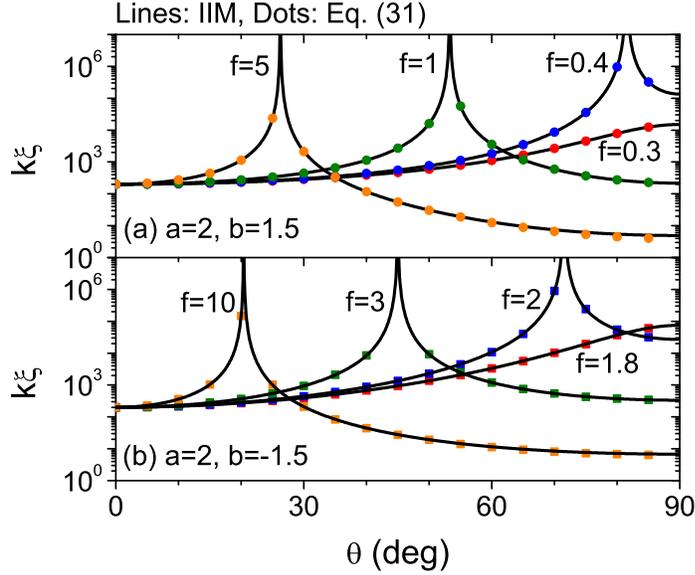}
\caption{Normalized localization length versus incident angle for $p$ waves in Model II, when $a=2$, $g_\perp=0.01$
and (a) $b=1.5$ and (b) $b=-1.5$, for designated values of $f$. The invariant imbedding results are compared with those obtained from Eq.~(\ref{eq:af2}).}
\label{fig3}
\end{figure}

In Fig.~3, we plot $k\xi$ versus $\theta$ for Model II, when
$a=2$, $b=\pm 1.5$ and $g_\perp=0.01$, for various values of $f$. The angle $\theta_B$ defined by Eq.~(\ref{eq:ba}) exists
when $0<b^2/(b+af)<1$ and $f>0$. In the case of Fig.~3(a) [3(b)], we get the BA if $f>0.375$ ($f>1.875$),
in a perfect agreement with the numerical results. The entire curves as well as the values of $\theta_B$ agree precisely with Eq.~(\ref{eq:af2}).
The BA has also been observed in Fig.~1(a) of Ref.~\cite{jor1}, where it is easy to see that the random variations of $\epsilon_\perp$
and $\epsilon_\parallel$ are directly proportional to each other with a positive ratio, in consistence with our results.

\begin{figure}[htbp]
\centering
\includegraphics[width=10cm]{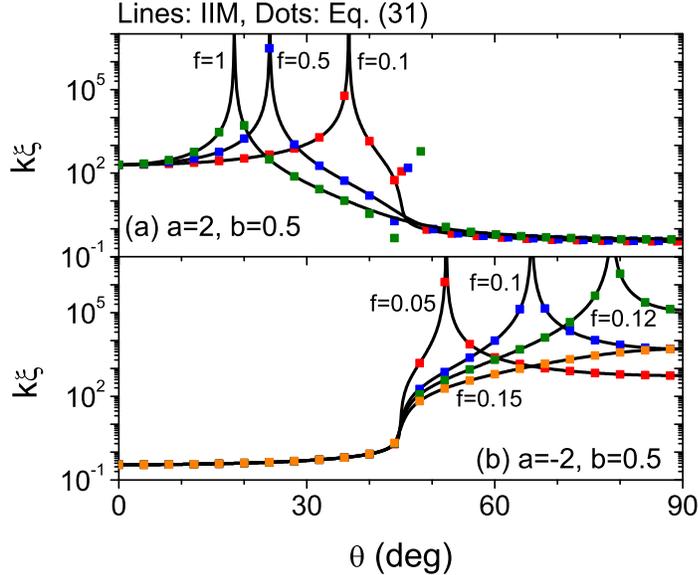}
\caption{Normalized localization length versus incident angle for $p$ waves in Model II, when $b=0.5$, $g_\perp=0.01$
and (a) $a=2$ and (b) $a=-2$, for designated values of $f$. The invariant imbedding results are compared with those obtained from Eq.~(\ref{eq:af2}). The discrepancy between the two is visible only near $\theta_c=45^\circ$ in (a).}
\label{fig4}
\end{figure}

Next, we consider the situation where $0<b<1$. There exists a critical angle of total reflection, $\theta_c$, given by
$\sin\theta_c=\sqrt{b}$. Then the BA can occur for both $a>1$ and $a<-1$ cases.
In Fig.~4, we plot $k\xi$ versus $\theta$ for Model II, when
$a=\pm 2$, $b=0.5$ and $g_\perp=0.01$, for various values of $f$. In the case of Fig.~4(a), the BA is possible for any value of $f>0$.
In the case of Fig.~4(b), it is possible only if $0<f<0.125$. We note that
$\theta_B<\theta_c=45^\circ$ if $a>0$, while $\theta_B>\theta_c$ if $a<0$. Interestingly, in the corresponding non-disordered case with $g_\perp=0$,
the ordinary Brewster angle, $\theta_b$, which is given by $\sin\theta_b=\left[b(1-a)/(1-ab)\right]^{1/2}$, exists only when $a$ is negative.
When $a=-2$ and $b=0.5$, $\theta_b$ is equal to $60^\circ$ ($>\theta_c$) and has no direct relationship to $\theta_B$.
When $a=2$ and $b=0.5$, no Brewster effect occurs in the clean case, still the BA can occur at an angle smaller than $\theta_c$
in the random case.

In Table 1, we make a comparison between the results of this work and those of Ref.~\cite{jor1}.
We remind that our model with $\delta$-correlated disorder is substantially different from the multilayer model of Ref.~\cite{jor1}
and only qualitative comparisons can be made. One of the biggest differences is that the ordinary Brewster angle
$\theta_b$ is the same as the angle $\theta_B$ where the BA would occur in Ref.~\cite{jor1}, while those two angles are unrelated in 
our work.

\begin{table}
\caption{Comparison between the results of this work and those of Ref.~\cite{jor1}.}
\begin{tabular*}{\textwidth}{@{}l*{15}{@{\extracolsep{0pt plus
12pt}}l}}
\br
&This work&Ref.~\cite{jor1}\\
\br
Relation between $\theta_b$ and $\theta_B$ & No relation & $\theta_b=\theta_B$\\
\mr
Model where the BA occurs & Model II with $f>0$   & Index-unmatched \\
&and weak disorder&binary stack\\
\mr
Nonrandom and matched $\epsilon_\parallel$,  & $\xi\propto (\cos\theta)^{-2}$ & $\xi\approx {\rm const.}$ for $\theta< 60^\circ$,\\ Random $\epsilon_\perp$&&$\xi\rightarrow\infty$ as $\theta\rightarrow 90^\circ$\\
\mr
Nonrandom $\epsilon_\perp$, Random $\epsilon_\parallel$ & $\xi\propto \theta^{-4}$ as $\theta\rightarrow 0$, &\\
&$\xi\rightarrow\infty$ as $g_\parallel\rightarrow \infty$& \\
\br
\end{tabular*}
\end{table}

\section{Conclusion}

In conclusion, we have studied Anderson localization and the BA of electromagnetic waves in random anisotropic media theoretically.
We have presented a unique perspective on the BA that
it is a phenomenon occurring
when the effective wave impedance is uniform and non-random, which is possible only in weakly-disordered media.
We have also argued that the Brewster effect occurs when the wave impedance is completely matched
throughout the space.
We have derived the existence condition for the BA and analytical expressions for the localization length and elucidated several
interesting physical aspects.
Our results can provide valuable insights in understanding the unique properties of some biological reflectors
and designing novel photonic devices based on anisotropic media \cite{jor3}.

\ack
This work has been supported by the National Research Foundation of Korea Grant (NRF-2018R1D1A1B07042629) funded by the Korean Government.

\end{document}